\documentstyle[11pt]{article}
\makeatletter
\@addtoreset{equation}{section}
\newdimen\normalarrayskip
\newdimen\minarrayskip
\normalarrayskip\baselineskip
\minarrayskip\jot
\newif\ifold \oldtrue \def\new{\oldfalse}
\def\arraymode{\ifold\relax\else\displaystyle\fi}

\def\@arrayskip{\ifold\baselineskip\z@\lineskip\z@
  \else
  \baselineskip\minarrayskip\lineskip2\minarrayskip\fi}
\def\@arrayclassz{\ifcase \@lastchclass \@acolampacol \or
\@ampacol \or \or \or \@addamp \or
 \@acolampacol \or \@firstampfalse \@acol \fi
\edef\@preamble{\@preamble
 \ifcase \@chnum
  \hfil$\relax\arraymode\@sharp$\hfil
  \or $\relax\arraymode\@sharp$\hfil
  \or \hfil$\relax\arraymode\@sharp$\fi}}
\def\@array[#1]#2{\setbox\@arstrutbox=\hbox{\vrule
  height\arraystretch \ht\strutbox
  depth\arraystretch \dp\strutbox
  width\z@}\@mkpream{#2}\edef\@preamble{\halign \noexpand\@halignto
\bgroup \tabskip\z@ \@arstrut \@preamble \tabskip\z@ \cr}%
\let\@startpbox\@@startpbox \let\@endpbox\@@endpbox
 \if #1t\vtop \else \if#1b\vbox \else \vcenter \fi\fi
 \bgroup \let\par\relax
 \let\@sharp##\let\protect\relax
 \@arrayskip\@preamble}
\@addtoreset{equation}{section}
\makeatother

\def\theequation{\thesection.\arabic{equation}}

\def\req#1{(\ref{#1})}

\def\commut#1#2{\left[{#1},\,{#2}\right]}

\def\const{\mathop{\rm const}\nolimits}

\def\bar{\overline}
\def\frac#1#2{\mathchoice{{\textstyle{{#1}\over{#2}}}}{{#1\over#2}}%
  {{#1\over#2}}{{#1\over#2}}}

\def\d{\partial}

\def\dl#1#2{{\stackrel{\rightarrow}{\d}\!#1\over\d #2}}

\def\ddl#1#2{{\stackrel{\rightarrow}{\d}\over\d #2}#1}
\def\ddr#1#2{#1{\stackrel{\leftarrow}{\d}\over\d #2}}

\def\half{\frac{1}{2}}

\def\sign#1#2{(-1)^{(\e{#1}+1)(\e{#2}+1)}}
\def\sig#1#2{(-1)^{\e{#1}\e{#2}}}

\def\cC{{\cal C}}

\def\cF{{\cal F}}

\def\cL{{\cal L}}
\def\cM{{\cal M}}
\def\cN{{\cal N}}

\def\CM{C_{\cM}}

\def\tilde{\widetilde}
\def\hat{\widehat}

\def\PRD{Phys.\ Rev.\ D}
\def\NPB{Nucl.\ Phys.\ B}
\def\PLB{Phys.\ Lett.\ B}
\def\MPLA{Mod.\ Phys.\ Lett.\ A}
\def\CMP{Commun.\ Math.\ Phys.}

\def\JMP{J.\ Math.\ Phys.}

\newtheorem{fact}{Proposition}[section]

\newtheorem{dfn}[fact]{Definition}

\newtheorem{thm}[fact]{Theorem}

\newcommand{\abrkt}[2]{({#1}\,,\,{#2})}

\newcommand{\vectorf}[1]{{\bf #1\,}}
\newcommand{\func}[1]{{{\cC}_{#1}}}

\newcommand\one{1\hspace{-4pt}1}
\newcommand{\e}[1]{\epsilon(#1)}

\def\Vect{{\rm Vect}}

\def\contraction{\vectorf i}
\def\I{{\bf I}}
\def\K{{\bf K}}
\def\J{{\bf J}}
\def\z{{z}}

\def\E{{\rm E}}

\begin{document}
\hfuzz=1pt
\addtolength{\baselineskip}{2pt}

\begin{flushright}
  {\tt hep-th/9807023}
\end{flushright}
\thispagestyle{empty}

\begin{center}
  {\Large{\sc A K\"ahler Structure of Triplectic Geometry}}\\[16pt]
  {\large M.~A.~Grigoriev and A.~M.~Semikhatov}\\[4pt]
  {\small\sl Tamm Theory Division, Lebedev Physics Institute, Russian
    Academy of Sciences}
\end{center}

\parbox{.9\textwidth}{\footnotesize We study the geometry of the
  triplectic quantization of gauge theories. We show that underlying
  the triplectic geometry is a K\"ahler manifold $\cN$ with a pair of
  transversal polarizations. The antibrackets can be brought to the
  canonical form if and only if $\cN$ admits a flat symmetric
  connection that is compatible with the complex structure and the
  polarizations.}

\section{Introduction}
The $Sp(2)$-symmetric Lagrangian quantization~\cite{[BLT],[Hull]} of
general gauge theories generalizes the standard
BV-formalism~\cite{[BV]} so that ghosts and antighosts enter it in a
symmetric way. The triplectic quantization~\cite{[BM0],[BMS],[BM]} has
been formulated as the corresponding analogue of the {\it covariant\/}
formulation of the BV scheme~\cite{[BT],[ASS0],[HZ],[SZ2],[ASS3]},
where ``covariant'' refers to the space of fields.  An essential point
in such a formulation is to ensure that the antibracket(s) can be
locally brought to the canonical (``Darboux'') form, since only then
the equivalence with the Hamiltonian quantization has been
established.

In this paper, we investigate the geometry underlying
the triplectic quantization procedure. There are
considerable differences from the usual BV
formalism. By construction, the covariant version of the BV scheme
does not differentiate between fields and antifields, which simply
become non-invariant notions.  In the triplectic formalism, on the
other hand, the antibrackets are degenerate, therefore one can single
out the {\it marked functions\/} (Casimir functions, or ``zero
modes'') of the antibrackets; the marked functions then span the space
of antifields.  In this sense, the antifields are already encoded in
the triplectic data.

As we will see, the triplectic geometry is
essentially concentrated on the
``manifold of antifields.''  This turns out to be a {\it complex\/}
manifold~$\cN$, the complex
structure originating from, and giving the geometric interpretation
of, the $e$-structure entering the weakly canonical antibrackets
from~\cite{[GS]}.  Further, the existence of {\it two\/} antibrackets
induces a polarization on~$\cN$, and the symmetrized Jacobi
identities~\cite{[BLT]} imply then that the associated Nijenhuis
tensor vanishes.  Finally, the one-form $\cF$
that enters the triplectic data~\cite{[BM],[GS]} (the ``potential''
for the odd vector fields) induces a {\it symplectic\/} structure on
$\cN$, which together with the complex structure makes it into a
K\"ahler manifold.

The properties of the ``antifield'' manifold $\cN$ are,
in particular, responsible for
the possibility of bringing the triplectic antibrackets to the
canonical form.
The condition for the general triplectic antibrackets to allow
the transformation
to the canonical form reformulates as the requirement that $\cN$
admit a flat symmetric connection that is compatible with the
complex structure and the polarization.  This solves the problem posed
in~\cite{[BMS]} and addressed recently in~\cite{[GS]}.

In Sec~\ref{sec:recall}, we briefly recall the triplectic formulation
and reformulate the structures known from~\cite{[GS]}.  In
Sec.~\ref{subsection:kahler}, we show how these translate into the
language of K\"ahler geometry. An important fact proved in
Sec.~\ref{sec:long} is that these geometric structures distinguish
different triplectic structures up to local equivalence. The geometric
reformulation, further, allows us to derive the conditions for the
existence of canonical coordinates for the antibrackets
(Sec.~\ref{subsec:finding}). In Sec~\ref{subsec:Sp2}, we briefly
discuss the $Sp(2)$ action, and in Sec.~\ref{sec:L}, we describe the
geometric restrictions arising on the manifold of {\it fields\/} of
the theory.

\section{Geometry of triplectic manifolds\label{sec:recall}}
\subsection{Basic definitions}
The geometric background of triplectic quantization is a
$(2N+2k|4N-2k)$-dimensional supermanifold $\cM$ endowed with a pair of
compatible antibrackets and an even 1-form~$\cF$, which we briefly
recall. Let $\func{\cM}$ be the algebra of smooth functions on $\cM$.
An antibracket $\abrkt{\;\cdot~}{\;\cdot~}$ is an odd skew-symmetric
bilinear map $\func\cM \times \func\cM \to \func\cM$ satisfying the
Leibnitz rule and Jacobi identity.  The triplectic antibrackets
$\abrkt{\;\cdot~}{\;\cdot~}^1$ and $\abrkt{\;\cdot~}{\;\cdot~}^2$ are
compatible in the following way:
\begin{equation}
  \label{Jacobi}
  (-1)^{(\e{F}+1)(\e{H}+1)}((F,G)^{\{a},H)^{b\}} +
  {\rm cycle}(F,G,H) =0\,,\quad
  F,G,H \in \func\cM\,,
\end{equation}
where the curly brackets stand for symmetrization of indices.  This
condition is often referred to as the symmetrized Jacobi identity
\cite{[BLT]}.  The antibrackets can be specified in terms of two
bivector fields $\E^a:\Omega_\cM\times\Omega_\cM\to\func\cM$
determined by $\E^a(dF,dG)=\abrkt{F}{G}$\,, with $d$ being the De Rham
differential of $\cM$ and $\Omega_\cM$ being the space of 1-forms
on~$\cM$ (while $\E^a(\phi_1,\phi_2)$ denotes the bivector $\E^a$
evaluated on the 1-forms $\phi_1$ and $\phi_2$).  In the local
coordinates $\z^A\,, A=1,\ldots,6N$ on $\cM$, we have
$\E^{aAB}=\E^a(d\z^A,d\z^B)=\abrkt{\z^A}{\z^B}^a$.

The bivector $\E^a$ determines a mapping from 1-forms into vector
fields that sends every 1-form $\phi$ to the vector field
$X^a=\E^a\phi$ such that $(\E^a\phi)G=\E^a(\phi,dG)$, $G\in\func\cM$.
In particular, the even 1-form $\cF$ gives rise to a pair of odd
vector fields $V^a=\E^a\cF$.  The triplectic quantization prescription
requires $V^a=\E^a\cF$ to be compatible with the antibrackets,
\begin{equation}
  V^{\{a}\abrkt{F}{G}^{b\}}-
  \abrkt{V^{\{a}F}{G}^{b\}}-(-1)^{\e{F}+1}
  \abrkt{F}{V^{\{a}G}^{b\}}=0\,,\quad F,G\in \func\cM\,,
\end{equation}
which can be rewritten as
\begin{equation}\label{dFcond}
  \Psi(\E^{\{a}\phi_1,\E^{b\}}\phi_2)=0\,,\quad \Psi=d\cF\,,
  \label{Psicond}
\end{equation}
for any 1-forms $\phi_1$ and $\phi_2$. (Here $\E^a\phi_1$ and
$\E^b\phi_2$ stand for the vector fields on which the 2-form $\Psi$ is
evaluated; see the Appendix for precise definitions of
differential-geometric objects).

In local coordinates $\z^A$, we write $\cF_A=\cF(\ddl{}{\z^A})$,
$\Psi_{AB}=\Psi(\ddl{}{\z^A},\ddl{}{\z^B})$. Then Eq.~\req{dFcond}
takes the form \cite{[BM]}
\begin{equation}
  \E^{ \{a AC } \Psi_{CD} \E^{b\} DB}=0 \,,
  \qquad \Psi_{AB}=\half(\d_A \cF_B -
  (-1)^{ \e{A}\e{B}}  \d_B \cF_A)(-1)^{\e{B}+1} \,.
  \label{coorddiff}
\end{equation}

\medskip

The additional constraints imposed on the triplectic data~\cite{[GS]}
are formulated in terms of the marked functions of the antibrackets.
A function $\varphi \in \CM$ is called a marked function of the
antibracket $(~,~)$ if $(F,\varphi)=0$ for any $F \in \func\cM$.  Two
compatible antibrackets $\abrkt{~}{~}^a$, $a=1,2$ are called {\it
  mutually commutative\/}\footnote{In~\cite{[GS]}, such antibrackets
  were called mutually flat; we now prefer changing the term, because
  in the geometric approach that we develop below, we do meet the
  actual flatness conditions; on the other hand, the condition in the
  text has the meaning of the {\it commutativity\/} of certain vector
  fields.} if any marked functions $\phi$ and $\psi$ of the first
antibracket $\abrkt{~}{~}^1$ satisfy $\abrkt{\phi}{\psi}^2=0$ and
conversely, the first antibracket vanishes when evaluated on marked
functions of the second antibracket.  A pair of antibrackets is called
{\it jointly nondegenerate\/} if the antibrackets do not have common
marked functions (i.e., bivectors $\E^1$ and $\E^2$ do not have common
zero modes).

We now introduce the notion of triplectic manifolds (see \cite{[GS]}
for the details).
\begin{dfn}\label{def:triplectic}
  A $(2N+2k|4N-2k)$-dimensional supermanifold $\cM$ endowed with a
  pair of compatible antibrackets and even 1-form $F$ is called {\em
    triplectic} if
  \begin{enumerate}\addtolength{\parskip}{-4pt}
  \item the antibrackets are jointly nondegenerate and mutually
    commutative,

  \item each of the antibrackets is of rank $4N$,

  \item the 2-form $\Psi=dF$ is compatible with the antibrackets (i.e.
    satisfies \req{Psicond}) and is of rank $4N$.
  \end{enumerate}
\end{dfn}
By the ranks of an antibracket and a 2-form $\Psi$, we mean the ranks
of the respective supermatrices $\E^{aAB}$, $a=1,2$, and $\Psi_{AB}$.

\subsection{Geometric objects on the triplectic manifold}
Let $\cM$ be a triplectic manifold.  In some neighbourhood $U$ of any
point of $\cM$ we can choose functions $\xi_{1i}\,,\xi_{2 \alpha}$,
$i,\alpha=1,\ldots,2N\,$ in such a way that $\xi_{1i}$ ($\xi_{2
  \alpha}$) is a minimal set that generates the algebra of marked
functions of the {\it second\/} (respectively, the {\it first\/})
antibracket.  We have shown in \cite{[GS]} that there exist functions
$x^i$ such that $(\xi_{1i},\xi_{2 \alpha},x^i)$ is a local coordinate
system in~$U$ in which the antibrackets take the form
\begin{equation}\label{weakcanonicalform}
  \new\begin{array}{rcl}
    (F,G)^1&=&\ddr{F}{x^i}\ddl{G}{\xi_{1i}}-
    \sign{F}{G}(F \leftrightarrow G)\,,\\
    (F,G)^2&=&\ddr{F}{x^i} \,e^i_\alpha\, \ddl{G}{\xi_{2 \alpha}}-
    \sign{F}{G}(F \leftrightarrow G)\,,
  \end{array}
\end{equation}
where $e^i_\alpha$ depend only on the marked functions $\xi_{1i}$ and
$\xi_{2 \alpha}$.  This form is called {\it weakly canonical}.  Now
the symmetrized Jacobi identity \req{Jacobi} rewrites as
\begin{equation}\new
  \begin{array}{rcl}
    \ddl{}{\xi_{1i}}e^j_\alpha-
    \sign{i}{j} \ddl{}{\xi_{1j}}e^i_\alpha&=&0\,,\\
    e^i_\alpha \ddl{}{\xi_{2\alpha}}e^j_\beta-
    \sign{i}{j} e^j_\alpha \ddl{}{\xi_{2\alpha}}e^i_\beta&=&0\,,
  \end{array}
  \label{econd}
\end{equation}
where we use the following Grassmann parity assignments:
$\e{x^i}=\e{i}\,,\e{\xi_{1i}}=\e{i}+1\,,
\e{\xi_{2\alpha}}=\e{\alpha}+1$. It also follows from the above rank
condition that $e^i_{\alpha}$ is an invertible matrix.

\smallskip

Each antibracket determines foliations $\cM_a \to \cM$, where $\cM_1$
($\cM_2$) is the symplectic leaf of the first (respectively, the
second) antibracket.  In the local coordinates $(\xi_{1i},\xi_{2
  \alpha},x^i)$, every submanifold $\cM_1$ ($\cM_2$) is singled out by
the equations $\xi_{2 \alpha}=\const_\alpha$ (respectively
$\xi_{1i}=\const_i$).  We also consider the foliation $i:\cL\to \cM$
with the fibres $\cL=\cM_1 \cap\cM_2$.

Using the weakly canonical coordinate system also allows us to
simplify the 2-form $\Psi=d \cF$. First of all we note that
compatibility condition \req{Psicond} implies that the 2-form $\Psi$
vanishes on a pair of vectors that are tangent to $\cM_1$ or $\cM_2$.
Condition \req{Psicond} also implies that the vectors tangent to $\cL$
are zero modes of~$\Psi$. Thus the only nonvanishing coefficients of
$\Psi$ are $\Psi^{i\alpha}=\sig{i}{\alpha} \Psi^{\alpha
  i}=\Psi(\ddl{}{\xi_{1i}},\ddl{}{\xi_{2 \alpha}})$,
\begin{equation}
  \Psi =2 d\xi_{1i} \wedge d\xi_{2 \alpha} \Psi^{\alpha i}\,.
  \label{Psiform}
\end{equation}
Since $\Psi$ is exact, the coefficients $\Psi^{i\alpha}$ are
independent of $x^i$.  In addition, the rank condition requires
$\Psi^{i\alpha}$ to be an invertible matrix.  Finally,
inserting~\req{Psiform} into~\req{Psicond}, we obtain the condition
\begin{equation}
  e^i_\alpha(\xi_1,\xi_2)\Psi^{\alpha j}
  +(-1)^{\epsilon(i)\epsilon(j)}
  e^j_\alpha(\xi_1,\xi_2)\Psi^{\alpha i}=0\,.
  \label{Tyutineq}
\end{equation}

An interesting feature of triplectic geometry is that the triplectic
data determine a Poisson bracket on the entire manifold~$\cM$.  This
originates from the bivector field \cite{[BM]} $\omega(\phi_1,\phi_2)=
-\half\epsilon_{ab}\Psi(\E^a\phi_1,\E^b\phi_2)$, which gives rise to
(see~\cite{[GS]} for the details)
\begin{equation}
  \{F,G\}=\omega(dF,dG)=F\ddr{}{\z^A} \omega^{AB} \ddl{}{\z^B}G\,,
  \label{PB}
\end{equation}
with $\omega^{AB}=\omega(d\z^A,d\z^B)$.  In the weakly-canonical
coordinates, the only nonvanishing coefficients of $\omega$ are
$\omega^{ij}=\omega(dx^i,dx^j)$, and therefore, the bracket \req{PB}
rewrites as
\begin{equation} \label{triPB}
  \{F,G\}= F\ddr{}{x^i} \omega^{ij} \ddl{}{x^j}G\,, \qquad
  \omega^{ij}=e^i_\alpha \Psi^{\alpha j}
  \,,
\end{equation}
where, moreover, $\omega^{ij}$ is an $x^i$-independent nondegenerate
matrix.  Thus the foliation into symplectic leaves of Poisson bracket
\req{triPB} coincides with the foliation $i:\cL \to \cM$ mentioned
above.  In particular, every leaf $\cL$ is a symplectic submanifold.

\subsection{The $\I$ structure}
The above structures defined on $\cM$ give rise to another structure
on the manifold.
\begin{fact}\label{fact:Iproperty}
  On a triplectic manifold $\cM$, there exists a tensor field
  $\I:\Vect_\cM \to \Vect_\cM$ {\rm(\/}and the transposed mapping
  $\I^T:\Omega_\cM \to \Omega_\cM$\/{\rm)} satisfying
  \begin{equation}\label{I2}
    \I^2=1\,,
  \end{equation}
  \begin{equation}\label{Iproperty}
    \E^1(\I^T \phi_1,\I^T
    \phi_2)=\E^2(\phi_1,\phi_2)
  \end{equation}
  for arbitrary 1-forms $\phi_1,\phi_2$, and
  \begin{equation}\label{Idef2}
    \I|_{T\cL}=\one\,.
  \end{equation}
  $\I$ acts on $d\xi_{1i},d\xi_{2 \alpha}$ as
  \begin{equation}
    \I^T d\xi_{2 \alpha}=d\xi_{1i} e^i_\alpha\,, \qquad
    \I^T d\xi_{1i}=d\xi_{2 \alpha} {\tilde e}^\alpha_i\,,
    \label{I-action-forms}
  \end{equation}
  where ${\tilde e}^\alpha_i$ is the inverse matrix to $e^i_\alpha$
  (i.e., $e^i_\alpha {\tilde e}^\alpha_j=\delta^i_j$).
\end{fact}

For a tensor field $\I : \Vect_\cM \to \Vect_\cM$, the transposed
mapping $\I^T:\, \Omega_\cM \to \Omega_\cM$ is defined by
\begin{equation}
  \langle X,\I^T\phi\rangle =\langle \I X,\phi\rangle \,,
\end{equation}
where $\langle X,\phi\rangle =\contraction_X\phi$ is the contraction
of the vector field $X$ with the 1-form $\phi$.  In the local
coordinates $\z^A$, we write $\I\ddl{}{\z^A}=\I_A^B\ddl{}{\z^B}$ and
$\I d\z^A =d\z^B\I_B^A$; then conditions~\req{I2} and \req{Iproperty}
become
\begin{equation}
  \I^C_B \I^A_C=\delta^A_B\,,\qquad
  (-1)^{\e{A}+\e{C}} \I^A_C \E^{1CD} \I_D^B=\E^{2AB}\,,
\end{equation}

The existence of a linear mapping satisfying \req{I2} and
\req{Iproperty} can be easily checked using the explicit form
\req{weakcanonicalform} of the antibrackets in the weakly canonical
coordinates.  Such a mapping is not unique.  However, {\it every $\I$
  satisfying \req{I2} and \req{Iproperty} can be restricted to the
  vector fields tangent to~$\cL$}.  Indeed, we can represent $X\in
\Vect_\cM$ that is tangent to $\cL$ as $X=\E^1\phi=\E^2\psi$ for some
$\phi,\psi \in \Omega_\cM$, because $T\cL=T{\cM_1} \cap T{\cM_2}$.
Then we have $\I X=\I (\E^1\phi)=\E^2 (\I^T\phi)$, which is thus
tangent to $\cM_1$; on the other hand $\I X=\I (\E^2\psi)=\E^1
(\I^T\psi)$ is tangent to $\cM_2$, which shows that $\I X$ is tangent
to $\cL$.

This allows us to impose condition \req{Idef2}.  Even this does not
completely fix the arbitrariness of~$\I$.  However, the action of $\I$
(in fact, of $\I^T$) on the 1-forms with vanishing restrictions to
$\cL$ is now unambiguous. In particular, $\I^T$ acts in a well-defined
way on the differentials $d\xi_{1i}$ and $d\xi_{2 \alpha}$.  In order
to find the explicit form of this action we consider the vector fields
\begin{equation}
  X^1_i=\abrkt{\xi_{1i}}{\cdot\,}^1=\E^1d\xi_{1i}\,,
  \qquad X^2_\alpha=\abrkt{\xi_{2 \alpha}}{\cdot\,}^2=\E^2d\xi_{2
    \alpha} \,. \qquad
\end{equation}
By definition, $X^1_i$ and $X^2_\alpha$ are tangent to $\cL$; in
addition, we have~\cite{[GS]} \ $X^2_\alpha=(-1)^{(\e{i}+1)\e{\alpha}}
e^i_\alpha X^1_i$, which we now rewrite as
\begin{equation}\label{XVfields}
  \I \E^1 (\I^Td\xi_{2 \alpha})=
  (-1)^{(\e{i}+1)\e{\alpha}} e^i_\alpha \E^1d\xi_{1 i}\,.
\end{equation}
Observing that $\I\E^2\I^Td\xi_{2\alpha}=0$ $\Rightarrow$ $\E^2\I^T
d\xi_{2\alpha}=0$ $\Rightarrow$ $\I^T
d\xi_{2\alpha}=d\xi_{1i}A^i_\alpha$ for some $A^i_\alpha$, we conclude
that $\E^1(\I^Td\xi_{2 \alpha})$ is tangent to $\cL$, which in turn
implies that $\E^1(\I^Td\xi_{2 \alpha})=\E^1(d\xi_{1i}e^i_\alpha)$.
Thus, $x\equiv\I^Td\xi_{2 \alpha}-d\xi_{1i}e^i_\alpha$ is a zero mode
of $\E^1$. Now, it is easy to see that $\E^2 x = 0$ as well, which
means that $x=0$ in view of the conditions imposed on the
antibrackets.  This shows~\req{I-action-forms}.

\section{From triplectic to K\"ahler geometry\label{subsection:kahler}}
As we have seen, a given triplectic structure determines a foliation
$i:\cL \to \cM$ of the triplectic manifold $\cM$ (the leaves being at
the same time the symplectic leaves of Poisson bracket~\req{triPB}).
For a sufficiently small neighbourhood $U$ in~$\cM$, this foliation is
a fibration with base~$U_\cN$ and the projection $\pi:U\to U_\cN$.
When the entire $\cM$ is a fibration, we will write $\pi:\cM\to\cN$,
then $U_\cN$ will be a neighbourhood in~$\cN$; however, it is not
necessarily assumed that $\cN$ exists globally, since we mainly work
with {\it local\/} statements. We identify the algebra $\func{U_\cN}$
of smooth functions on $U_\cN$ with the functions on $U$ that are
constant along the fibres; this gives precisely the algebra generated
by the marked functions of the antibrackets in the neighbourhood.
Further, the weakly canonical coordinates provide us with a
diffeomorphism $\psi:U\to U_\cN\times U_\cL$ that identifies $U$ with
a neighbourhood in the product of linear (super)spaces $U_\cN\times
U_\cL$ (such that the first component of $\psi$ is $\pi$, i.e.,
$\psi=(\pi,\rho):U\to U_\cN\times U_\cL$).

In the present section, we assume for simplicity that the base $\cN$
and the projection $\pi:\cM\to\cN$ exist globally.  Thus, smooth
functions on $\cN$ can be identified with functions on $\cM$ that are
constant along the fibres, i.e., with the algebra generated by the
marked functions of the antibrackets on~$\cM$.

In particular, we can choose a coordinate system
${\hat\xi}_{1i},{\hat\xi}_{2\alpha}$ on $\cN$ such that the functions
$\xi_{1i}=\pi^* {\hat\xi}_{1i}$ (respectively,
$\xi_{2\alpha}=\pi^*{\hat\xi}_{2\alpha}$), where $\pi^*$ is the
pullback associated with the projection $\pi$, generate the algebra of
marked functions of the second (respectively, the first) antibracket.
In what follows, we will not write the tilde over the coordinates on
$\cN$ and thus identify functions on $\cN$ with their pullbacks
to~$\cM$, in accordance with the one-to-one correspondence between
functions from $\func\cN$ and the functions that are constant
along~$\cL$.

Further, since $e^i_\alpha$ are constant along $\cL$, the 1-forms
$d\xi_{1i}e^i_{\alpha}$ and $d\xi_{2 \alpha}{\tilde e}^{\alpha}_j$ are
the pullbacks of some 1-forms on $\cN$ (as, obviously, are the 1-forms
$d\xi_{1i}$ and $d\xi_{2 \alpha}$).  Then, according to
proposition~\ref{fact:Iproperty}, we conclude that
$\I^T:\Omega_\cM\to\Omega_\cM$ determines a mapping
${\hat\I}^T:\Omega_\cN\to\Omega_\cN$, and thus ${\hat\I}$ is
well-defined on $\cN$.  In the local coordinates $\xi_{1i},\xi_{2
  \alpha}$ on $\cN$, we have
\begin{equation}
  {\hat\I}\ddl{}{\xi_{1i}}=e^i_\alpha\ddl{}{\xi_{2 \alpha}}\,, \qquad
  {\hat \I}\ddl{}{\xi_{2\alpha}}={\hat e}^\alpha_i\ddl{}{\xi_{1
      i}}\,.
  \label{I-action-vectors}
\end{equation}

Further, it follows from \req{Psiform} that there exists a 2-form
${\hat \Psi}$ on $\cN$ whose pullback coincides with $\Psi=d\cF$ from
\req{Psicond}.  The rank assumption implies that ${\hat \Psi}$ is
nondegenerate and, thus, $\cN$ {\it is a symplectic
  manifold}.\footnote{Note that the 2-form ${\hat \Psi}$ is in general
  closed but not exact, whereas $\Psi=d\cF$ is evidently exact.}  As
can be seen from \req{Psiform} and \req{Tyutineq}, the structures
identified on~$\cN$ are related by
\begin{equation}
  {\hat\Psi}({\hat\I}Y_1,{\hat\I}Y_2)=-{\hat\Psi}(Y_1,Y_2)
  \label{IPsi}
\end{equation}
for arbitrary vector fields $Y_1,Y_2$ on $\cN$.

\medskip

As regards vector fields, we have, obviously,
\begin{fact}\label{fact:VFproj}
  Every vector field $X:\func\cM \to \func\cM$ preserving the space of
  functions that are constant along $\cL$, determines a unique vector
  field $\hat X$ on $\cN$.
\end{fact}

We now show that the symplectic manifold $\cN$ is endowed with a pair
of transversal polarizations. We first recall that an integrable
distribution $P:\cN \to T\cN$ is called a {\it polarization\/} of the
symplectic manifold $\cN$ if the image $P_x \subset T_x \cN$ at any
point $x \in \cN$ is a Lagrangian subspace of~$T_x \cN$.  Two
polarizations $P^1$ and $P^2$ are called {\it transversal\/} if $T_x
\cN=P^1_x \oplus P^2_x$.

In the case at hand, we observe that the vector fields on $\cN$
annihilating the marked functions of the first antibracket (of the
second antibracket) considered as functions on $\cN$ determine a
foliation of $\cN$ and, thus, an integrable distribution $P^1:\cN \to
T\cN$ (respectively, $P^2:\cN \to T\cN$).  In the coordinate system
$\xi_{1i}\,,\xi_{2 \alpha}$ on $\cN$, we see that $P^1$ (respectively,
$P^2$) is generated by the vector fields $\ddl{}{\xi_{1i}}$
(respectively, $\ddl{}{\xi_{2 \alpha}}$).  The explicit form of $P^1$
and $P^2$ shows that $T_x \cN=P^1_x \oplus P^2_x$ at any point $x\in
\cN$.  It is easy to see that the symplectic form $\hat\Psi$ vanishes
on $P^1_x$ as well as on $P^2_x$, and, thus, $P^1$ and $P^2$ are a
pair of transversal polarizations.

Now, one can represent any vector field $X$ on $\cN$ as a sum
$X=X^1+X^2$, where $X^1 \in P^1$ and $X^2 \in P^2$.  This allows us to
introduce the mapping $\K:\Vect_\cN \to \Vect_\cN$ as
\begin{equation}
  \K X=X_1-X_2\,.
  \label{Kdef}
\end{equation}
It is easy to see that $\K$ satisfies
\begin{equation}
  \K^2=\K \K=\one \,,\quad
  \K{\hat \I}+{\hat \I}\K=0 \,,\quad {\hat\Psi}(\K Y_1,\K
  Y_2)=-{\hat\Psi}(Y_1,Y_2)\,,\quad Y_1,Y_2\in\Vect_\cN\,.
  \label{Kproperty}
\end{equation}

Given the mappings ${\hat\I}$ and $\K$ on $\cN$, we can consider the
product $\J={\hat\I}K$.  It follows from~\req{IPsi}
and~\req{Kproperty} that $\J$ satisfies
\begin{equation} \J^2=\J\J=-\one\,, \qquad
  {\hat\Psi}({\J}Y_1,{\J}Y_2)={\hat\Psi}(Y_1,Y_2)\,.
\end{equation}
Thus $\J$ is an almost complex structure which is compatible with the
symplectic form.  In the local coordinates $\xi_{1i},\xi_{2\alpha}$,
we have
\begin{equation}
  \J\ddl{}{\xi_{1i}}=e^i_\alpha\ddl{}{\xi_{2 \alpha}}\,, \qquad
  \J\ddl{}{\xi_{2\alpha}}=-{\tilde e}^\alpha_i\ddl{}{\xi_{1 i}}\,.
  \label{Jaction}
\end{equation}

Next, we show that this almost complex structure is integrable.  For
$\J$ to be integrable it is sufficient that the Nijenhuis tensor
$N_{\J,\J}$ vanish. The Nijenhuis tensor of $J$ is the mapping
$N_{\J,\J}: \Vect_\cN \times \Vect_\cN \to \Vect_\cN$ given by
\begin{equation}
  N_{\J,\J}(X,Y)=\commut{\J X}{\J
    Y}+\J \J \commut{X}{Y} -\J \commut{X}{\J Y} -\J\commut{\J
    X}{Y}\,,\quad X,Y \in \Vect_\cN\,.
\end{equation}
Using the explicit form of $\J$ given in \req{Jaction} we conclude
that $N_{\J,\J}=0$ in view of Eqs.~\req{econd}.  Therefore, {\it $\cN$
  is a complex manifold}.

Putting everything together, we have
\begin{thm}
  The manifold $\cN$ is K\"ahler.  The corresponding fundamental
  2-form is $\hat\Psi$.
\end{thm}
Since $\cN$ is in general a supermanifold, we actually have the super
analogue of a K\"ahler manifold.  Also, we have not required $h$ to be
positive definite, which means that $\cN$ is in fact a {\it
  pseudo\/}-K\"ahler manifold.

Explicitly, the K\"ahler metric is $h(X,Y)={\hat\Psi}(\J X,
Y)(-1)^{\e{Y}}$, $X,Y \in \Vect_\cN$.  It follows from the above that
$h$ is nondegenerate and satisfies
\begin{equation}
  h(X,Y)=\sig{X}{Y}h(Y,X)\,,\quad h(\J X,\J Y)=h(X,Y) \,,\quad X,Y \in
  \Vect_\cN\,.
\end{equation}

\section{Local equivalence of triplectic  manifolds\label{sec:long}}
We show in Sec~\ref{subsec:equivalence} that the geometric structures
induced on $U_\cN$ (see the beginning of Sec.~\ref{subsection:kahler})
distinguish different triplectic structures up to local equivalence.
In particular, the condition for the triplectic antibrackets to admit
the canonical form also reformulates in terms of some objects
on~$U_\cN$, as we show in Sec.~\ref{subsec:finding}.

\subsection{The equivalence theorem\label{subsec:equivalence}}
For a sufficiently small neighbourhood $U\subset\cM$, the triplectic
data give rise to the projection $\pi:U\to U_\cN$ along the leaves
$\cL$ of the foliation $i : \cL \to \cM$. The triplectic antibrackets,
further, induce a complex structure and a pair of transversal
polarizations on~$U_\cN$. Similarly, the 2-form $\Psi =d \cF$
determines the fundamental form of $U_\cN$.

We will say that two pairs of triplectic antibrackets\footnote{For
  brevity, we consider only the antibrackets, disregarding the odd
  nilpotent vector fields (i.e., the 1-form~$\cF$); the equivalence
  statement given below can easily be generalized to include~$\cF$.}
$(~,~)^a$ and ${\bar{(~,~)}}^a$ are locally equivalent if for any
sufficiently small neighbourhood $U\subset\cM$ there exists a
diffeomorphism $\phi:U\to U$ such that
\begin{equation}
  {\bar{(F,G)}}^a=(\phi^{-1})^*( \phi^*
  F,\phi^* G)^a \,,
  \label{ABequiv}
\end{equation}

Let now $(~,~)^a$ and ${\bar{(~,~)}}^a$ be two triplectic structures
on~$\cM$.  We show that different triplectic structures are locally
distinguished by geometries on~$U_\cN$.
\begin{thm}
  The following conditions are equivalent:
  \begin{enumerate}\addtolength{\parskip}{-6pt}
  \item The triplectic structures $(~,~)^a$ and ${\bar{(~,~)}}^a$ are
    locally equivalent.

  \item For every sufficiently small neighbourhood $U\subset\cM$,
    there exists a diffeomorphism $\phi_0 : U_\cN \to {\bar U}_\cN$
    such that
    \begin{equation}
      \phi_0^*({\bar \K}\psi)= \K ( \phi_0^* \psi )\,,\qquad
      \qquad \phi^* ({\bar \J} \psi)= \J (\phi_0^* \psi)\,,
      \label{Nequiv}
    \end{equation}
    for arbitrary 1-form $\psi$ on ${\bar U}_\cN$, where $\pi:U\to
    U_\cN$ and $\bar\pi:U\to{\bar U}_\cN$ are the projections
    associated with the respective triplectic structures.
  \end{enumerate}
\end{thm}

To show this, let $\phi: U\to U$ be a diffeomorphism satisfying
\req{ABequiv}.  Let also $\xi_{1i}$ ($\theta_{1i}$) be the marked
functions of the $\abrkt{\cdot}{\cdot}^2$ antibracket (respectively,
of ${\bar{\abrkt{\cdot}{\cdot}}}^2$) and $\xi_{2\alpha}$
($\theta_{2\alpha}$) be the marked functions of
$\abrkt{\cdot}{\cdot}^1$ (respectively, of
${\bar{\abrkt{\cdot}{\cdot}}}^1$).  It follows from \req{ABequiv} that
$\phi^*\theta_{1i}$ and $\phi^*\theta_{2 \alpha}$ are marked functions
of the brackets $\abrkt{\cdot}{\cdot}^2$ and $\abrkt{\cdot}{\cdot}^1$,
respectively, and therefore, $\phi^*\theta_{1i}$ is a function of
$\xi_{1i}$, which we write as
$\phi^*\theta_{1i}={\bar\xi}_{1i}(\xi_1)$ and similarly, $\phi^*
\theta_{2 \alpha}={\bar\xi}_{2 \alpha}(\xi_2)$.  Thus $\phi$ induces a
mapping from the marked functions of the
${\bar{\abrkt{\cdot}{\cdot}}}^a$ antibrackets to the marked functions
of the $\abrkt{\cdot}{\cdot}^a$ antibrackets.  We now consider the
vector fields generated by the marked functions
\begin{equation}
  (\xi_{2 \alpha},\cdot ~)^2 =-(-1)^{
    (\e{i}+1) \e{\alpha} } e^i_\alpha (\xi_{1i},\cdot ~)^1\,,\qquad
  {\bar{(\theta_{2 \alpha},\cdot~})}^2 =-(-1)^{ (\e{i}+1)\e{\alpha} }
  {\bar e}^i_\alpha {\bar{(\theta_{1i},\cdot~)}}^1\,,
  \label{ebare}
\end{equation}
where $e$ and $\bar e$ are the corresponding $e$-structures.
According to \req{ABequiv}, we have
\begin{equation}
  \abrkt{\phi^* \theta_{2 \alpha}}{\cdot\,}^2=
  -(-1)^{(\e{\i}+1)\e{\alpha}} (\phi^* {\bar e}^i_\alpha)
  \abrkt{\phi^* \theta_{1i}}{\cdot\,}^1\,.
\end{equation}
Since, as we have seen, $\phi^*\theta_{1i}={\bar\xi}_{1i}(\xi_1)$
and $\phi^*\theta_{2\alpha}={\bar\xi}_{2\alpha}(\xi_2)$, we have
\begin{equation}
  e^i_\alpha=\dl{{\bar \xi}_{1j}}{\xi_{1i}} (\phi^* { \bar e}^j_\beta)
  \dl{\xi_{2_\alpha}}{{\bar \xi}_{2 \beta}}\,.
  \label{ediff}
\end{equation}
Taking the marked functions $\xi_1,\,\xi_2$ as the coordinates on
$U_\cN$ and, similarly, $\theta_1,\,\theta_2$ as the coordinates on
${\bar U}_\cN$, we see that $\phi$ restricts to a diffeomorphism
$\phi_0:U_\cN\to{\bar U}_\cN$.  Recalling that $\phi^*$ maps marked
functions into the corresponding marked functions and also using
Eq.~\req{ediff}, we conclude that $\phi_0$ is as required in the
theorem.

\smallskip

Conversely, let $U_\cN$ and ${\bar U}_\cN$ be related by a
diffeomorphism $\phi_0$ satisfying~\req{Nequiv}.  We then choose a
coordinate system $\xi_{1i},\xi_{2 \alpha}$ (a coordinate system
$\theta_{1i},\theta_{2 \alpha}$) on $U_{\cN}$ (respectively, on ${\bar
  U}_\cN$) such that $\K$ and $\J$ (respectively, $\bar \K$ and $\bar
\J$) act on the basis 1-forms as \ $\K^T d\xi_{1i}=d\xi_{1i}$, $\K^T
d\xi_{2 \alpha}=-d\xi_{2\alpha}$ and $\J^T d\xi_{1i}=-{\tilde
  e}_i^\alpha d\xi_{2 \alpha}$, $\J^T d\xi_{2 \alpha}=e^i_\alpha
d\xi_{1i}$ (respectively, ${\bar \K}^T d\theta_{1i}=d\theta_{1i}$,
${\bar \K}^T d\theta_{2 \alpha}=-d\theta_{2\alpha}$ and ${\bar \J}^T
d\theta_{1i}=-{\bar {\tilde e}}_i^\alpha d\theta_{2 \alpha}$, ${\bar
  \J}^T d\theta_{2 \alpha}={\bar e}^i_\alpha d\theta_{1i}$).  Then
Eqs.~\req{Nequiv} imply that $\phi_0^* d \theta_{1i}=A_i^j d
\xi_{1j}$.  This, in turn, shows that $\phi_0^* \theta_{1i}$ are
functions of only $\xi_1$.  Similarly, $\phi_0^* \theta_{2 \alpha}$ is
a function of only $\xi_2$. This allows us to choose coordinates
${\bar \theta}_{1i}\,,{\bar \theta}_{2 \alpha}$ in ${\bar U}_n$ such
that $\phi_0^* {\bar \theta}_{1i}=\xi_{1i}$ and $\phi_0^*
{\bar\theta}_{2 \alpha}=\xi_{2 \alpha}$.  In coordinates, we write
${\bar \J}^T d{\bar \theta}_{2 \alpha} =({\bar e^\prime})^i_\alpha
d{\bar \theta}_{1i})\,$.  Then the second equation in~\req{Nequiv}
implies
\begin{equation}
  e^i_\alpha=\phi_0^*(({\bar e^\prime})^i_\alpha)\,,
\end{equation}
where we view $({\bar e^\prime})^i_\alpha$, for each $i$ and $\alpha$,
as functions on ${\bar U}_\cN$.  Further, we consider
$\xi_{1i},\,\xi_{2 \alpha}$ and
${\bar\theta}_{1i},\,{\bar\theta}_{2\alpha}$ as functions on $U
\subset\cM$, where they are the marked functions of corresponding
antibrackets.  Choosing the functions $x^i$ and $y^i$ on $U\subset\cM$
in such a way that $x^i,\,\xi_{1i},\,\xi_{2\alpha}$ (respectively,
$y^i,\,{\bar\theta}_{1i},\,{\bar\theta}_{2\alpha}$) be the weakly
canonical coordinates for the antibrackets $(\cdot,\cdot)^a$
(respectively, ${\bar{(\cdot,\cdot)}}^a$), we consider the
diffeomorphism $\phi:U\to U$ determined by
\begin{equation}
  \phi^* y^i=x^i\,,\quad \phi^*{\bar
    \theta}_{1i}=\xi_{1i} \,, \quad \phi^* {\bar \theta}_{2
    \alpha}=\xi_{2 \alpha}\,.
\end{equation}
It is easy to check that $\phi$ satisfies \req{ABequiv} and, thus, two
triplectic structures are locally equivalent.  This completes the
proof.

\subsection{Finding the canonical coordinates\label{subsec:finding}}
As we are going to see, the question of whether the antibrackets can
be (locally) brought to the canonical form is solved in terms of
geometric structures on~$U_\cN$. Recall that having chosen the bases
of marked functions of the antibrackets, one arrives at the structure
$e^i_\alpha(\xi_{1},\xi_{2})$, which is in general a local obstruction
to finding the canonical coordinates for the triplectic antibrackets.
It follows from \req{XVfields} that if we choose new bases of the
marked functions as $\xi^{\prime}_{1i}=\xi^{\prime}_{1i}(\xi_{1})$ and
$\xi^\prime_{2\alpha}=\xi^\prime_{2\alpha}(\xi_{2})$, the matrix $e$
transforms as follows:
\begin{equation}
  {e^{\prime}}^i_\alpha= \dl{ \xi_{1 j} }{ \xi^{\prime}_{1 i} }
  \,e^j_\beta\, \dl{ \xi^{\prime}_{2 \alpha} }{ \xi_{2 \beta} }\,,
  \qquad
  \abrkt{\xi^\prime_{2 \alpha}}{\;\cdot\;}^2 = (-1)^{
    (\e{i}+1)\e{\alpha}} {e^{\prime}} ^i_\alpha
  \abrkt{\xi^{\prime}_{1i}}{\;\cdot\;}^1 \,.
  \label{etransform}
\end{equation}
The structure $e$ is called {\it reducible} if there exist bases of
the marked functions $\xi^{\prime}_{1i}=\xi^\prime_{1i}(\xi_{1})$ and
$\xi^{\prime}_{2\alpha}=\xi^{\prime}_{2\alpha}(\xi_{2})$ such that
${e^\prime}^i_\alpha=\delta^i_\alpha$. Once $e$ is reducible, there
exists a coordinate system where both antibrackets take the canonical
(``Darboux'') form.

We now reformulate the problem of reducibility in terms of
differential geometry on $U_\cN$ from the previous section (the proof
of the following proposition is immediate from the explicit form of
$\J$, $\K$ and $\I=\J\K$, where we remove the hat over $\I$).
\begin{fact}
  Let $U$ be a (sufficiently small) neighbourhood in~$\cM$ and
  $\pi:U\to U_\cN$ be the projection associated with the triplectic
  structure. Then the following conditions are equivalent:
  \begin{enumerate}\addtolength{\parskip}{-6pt}

  \item The $e$-structure associated with the antibrackets is
    reducible.

  \item\label{item:constant} There exists a coordinate system in
    $U_\cN$, where the components of the tensor fields $\I$, $\K$, and
    $\J=\I\K$ are constants.

  \item\label{item:connection} There exists a flat symmetric linear
    connection $\nabla$ on $U_\cN$ such that the tensor fields $\I$,
    $\K$, and $\J$ are parallel with respect to~$\nabla$,
    \begin{equation}
      \nabla \K=0\,,\qquad \nabla \I=0\,,\qquad \nabla \J=0\,.
      \label{covconst}
    \end{equation}
  \end{enumerate}
\end{fact}
In items~\ref{item:constant} and~\ref{item:connection}, it suffices to
have the conditions satisfied for any two structures of $\I$, $\K$,
and $\J$. The covariant derivative $\nabla$ is viewed as a mapping
$\nabla:\Vect_{U_\cN}\times \Vect_{U_\cN}\to \Vect_{U_\cN}$ satisfying
\begin{equation}
  \nabla_{FX} Y=F\nabla_{X}
  Y\,,\quad
  \nabla_{X}(FY)=(XF)(Y)+\sig{X}{F}F\nabla_{X}Y\,,
\end{equation}
where $F\in\func{U_\cN}\,$ and $X,Y \in \Vect_{U_\cN}\,.$ The action
of $\nabla$ on a tensor field $I:\Vect_{U_\cN}\to\Vect_{U_\cN}$ is
defined by
\begin{equation}
  (\nabla_X I)Y=\nabla_X (I Y) - I(\nabla_X Y ) \,,
  \quad X,Y\in\Vect_{U_\cN}\,.
\end{equation}
Taking $\nabla$ symmetric means the vanishing of torsion
$T(X,Y)=\nabla_X Y-\sig{X}{Y}\nabla_Y X-\commut{X}{Y} $.  With the
Christoffel symbols defined in local coordinates $\z^A$ on ${U_\cN}$
as $\Gamma^C_{AB}=(\nabla_A\ddl{}{\z^B})\z^C$ and $I\ddl{}{\z^A}=I^B_A
\ddl{}{\z^B}$, we have
\begin{equation}
  (\nabla_A I)_B^C=\d_A I_B^C -\Gamma^D_{AB}
  I^C_D+ (-1)^{(\e{D}+\e{B})\e{A}} I^D_B\Gamma^C_{AD}\,.
\end{equation}

We further observe that the flat connection from
item~\ref{item:connection} is {\it unique}.  Indeed, it follows from
the first equation in~\req{covconst} and definition~\req{Kdef} of $\K$
that the only nonvanishing connection coefficients in the coordinates
$\xi_{1i},\xi_{2 \alpha}$ are $\Gamma^{ij}_k$ and $\Gamma^{\alpha
  \beta}_\gamma$.  Then, we use the equation $\nabla\I=0$ (and the
explicit form \req{I-action-vectors} of $\I$) to obtain
\begin{equation}
  \Gamma^{ij}_{k}
  =( \ddl{}{\xi_{1i}} e^j_\alpha )
  {\tilde e}^\alpha_k \,,\qquad
  \Gamma^{\alpha \beta}_\gamma=
  (\ddl{}{\xi_{2 \alpha}}{\tilde e}^\beta_j )
  e^j_\gamma\,,
  \label{connection}
\end{equation}
which shows that the symmetric connection $\nabla$ satisfying
$\nabla\I=\nabla\J= \nabla\K=0$ is unique.

Looking at the zero-curvature conditions, we see that the only
nonvanishing curvature components are
$\commut{\nabla^{1i}}{\nabla^{2\alpha}}$, therefore the flatness
condition becomes
\begin{equation}
  \ddl{}{\xi_{1i}}\biggl(\Bigl(\ddl{}{\xi_{2
      \alpha}}{\tilde e}^\beta_j\Bigr)\, e^j_\gamma\biggr) =0\,,
  \qquad
  \ddl{}{\xi_{2\alpha}}\biggl(\Bigl(\ddl{}{\xi_{1i}}e_\beta^j
  \Bigr)\,{\tilde e}_k^\beta\biggr) =0\,.
  \label{flat}
\end{equation}

We now recall that this vanishing curvature condition on ${U_\cN}$ can be
traced back to the reducibility of $e^i_\alpha$ on $\cM$.  This gives
the following theorem on the transformation of the triplectic
antibrackets to the canonical form.
\begin{thm}
  The structure $e^i_\alpha$ corresponding to the triplectic
  antibrackets is reducible if and only if it satisfies \req{flat}. \
  Thus, the triplectic antibrackets (i.e., a pair of rank-$4N$
  compatible antibrackets that are jointly nondegenerate and mutually
  commutative) admit canonical coordinates if and only if the
  corresponding $e$-structure satisfies~\req{flat}.
\end{thm}

\section{The $Sp(2)$ action on $\cN$\label{subsec:Sp2}}
In this section, we return, for simplicity, to the situation described
in Sec.~\ref{subsection:kahler}, where the base $\cN$ is assumed to
exist globally.

An essential ingredient of the ghost-antighost symmetric quantization
is the $Sp(2)$ action~\cite{[BLT]}.  In the covariant formulation,
this takes the form of the requirement that $\cM$ should carry an
action of~$Sp(2)$~\cite{[GS]}, i.e., for every $G \in Sp(2)$ there is
a mapping $\phi_G : \cM \to \cM$ such that
$\phi_{G_1}\phi_{G_2}=\phi_{G_1 G_2}$.  A pair of antibrackets and a
1-form $\cF$ are called $Sp(2)$-covariant if there exists an action
$\phi$ of $Sp(2)$ on~$\cM$ such that
\begin{equation}
  \phi^*_G((f,g)^a)=
  G^a_b(\phi^*_G(f),\phi^*_G(g))^b\,,
  \quad \phi^*_G\cF=\cF\,,\qquad G\in Sp(2)\,,\quad
  f,g\in C_{\cM}\,,
  \label{spcovariance}
\end{equation}
where $G^a_b$ is the $2\times2$ matrix representation of $Sp(2)$.
Infinitesimally, this reformulates as a homomorphism from the Lie
algebra $sp(2)$ to $\Vect_\cM$ such that
\begin{equation}\label{Ecov}
  L_Y \E^a =g^a_b E^{b}\,,\qquad
  L_Y \cF  =0\,,
\end{equation}
where $Y$ is the vector field corresponding to $g\in sp(2)$ (and $L$
is the Lie derivative). This has an important consequence that Poisson
bracket~\req{triPB} is $sp(2)$-invariant:
\begin{equation}
  L_Y\omega = 0\,.
\end{equation}

Next, we observe that the $Sp(2)$ action maps the marked functions
$(\xi_{1i},\xi_{2\alpha})$ into marked functions (but does not,
obviously, preserve the separation of the marked functions into those
of the first and the second antibracket). A convenient way to see this
is to note that the collection $(\xi_a)=(\xi_{1i},\xi_{2\alpha})$ of
marked functions can be characterized by the fact that these are
marked functions of the Poisson bracket, $\{F,\xi_a\}=0$ for any $F$.
Applying now an $Sp(2)$ transformation, we have
\begin{equation}
  0=\phi_G^*(\{F,\xi\})=\{\phi_G^*(F),\phi_G^*(\xi)\}\,,
\end{equation}
which means that $\phi_G^*(\xi)$ is again a marked function of the
Poisson bracket. This implies, further, that the vector fields
representing the $sp(2)$ action on $\cM$ project onto
$\cN$;\footnote{As noted above, we now assume that $\cN$ exists
  globally.} we will denote the projection of $Y$ again by~$Y$.  Since
$L_Y\cF$ = 0 and hence $L_Y\Psi=0$, we arrive at\footnote{Although the
  shortest way to this statement is to recall the Poisson bracket, it
  can also be shown without resorting to the Poisson structure.}
\begin{fact}
  The manifold $\cN$ carries an $Sp(2)$ action that preserves the
  symplectic structure on~$\cN$.
\end{fact}
Thus, the vector fields $Y$ are locally Hamiltonian on $\cN$.

Choosing now $Y^\pm$ and $Y^0$ to correspond to the basis in $sp(2)$
where
\begin{equation}
  \commut{J^+}{J^-}=-2J^{0}  \,, \qquad \commut{J^+}{J^0}=-J^{+}   \,,
  \qquad \commut{J^{-}}{J^{0}}=J^{-}\,,
\end{equation}
we see that the structures $\I$, $\J$, and $\K$ furnish the
three-dimensional representation of $sp(2)$:
\begin{equation}
  \label{Yaction}
  \new\begin{array}{ccc}
    L_{Y^+}\I=\K\,,&\qquad L_{Y^-}\I=-\K\,,&\qquad L_{Y^0}\I=\J\,, \\
    L_{Y^+}\J=-\K\,,&\qquad L_{Y^-}\J=-\K\,,&\qquad L_{Y^0}\J=\I\,, \\
    L_{Y^+}\K=-\I-\J\,,&\qquad L_{Y^-}\K=\I-\J\,,&\qquad L_{Y^0}\K=0\,.
  \end{array}
\end{equation}

Apart from the {\it global\/} properties of the group action on a
manifold, the issue of $Sp(2)$ covariance of triplectic antibrackets
is solved for the entire class of equivalent triplectic structures
and, therefore, can be solved in terms of geometry on~$\cN$---it
amounts to the existence of an $Sp(2)$ action on~$\cN$
satisfying~\req{Yaction}.

\section{Geometry of $\cL$\label{sec:L}}
In this section, we will show that a pair of antibrackets induce an
additional structure on every submanifold $\cL$ (every leaf of the
foliation $i:\cL\to\cM$). Besides the known symplectic structure
on~$\cL$, the conditions imposed on the triplectic objects (see
Definition~\ref{def:triplectic}) imply the existence of a flat
connection on~$\cL$:
\begin{thm}
  A pair of triplectic antibrackets induce a flat symmetric connection
  on each leaf $\cL$ of the foliation $i:\cL\to \cM$.
\end{thm}

To prove the theorem, we choose a fixed leaf $\cL\subset\cM$; let
$\{U_n\}$ be an atlas of $\cM$ such that in each neighbourhood $U_n$
there exist weakly canonical coordinates $x^i,\xi_{1i},\xi_{2
  \alpha}$.  Let $U_1$ and $U_2$ be coordinate neighbourhoods on $\cM$
such that ${\bar U}_1=U_1 \cap \cL$ and ${\bar U}_2=U_2 \cap \cL$, and
also ${\bar U}_1\cap {\bar U}_2$, are non-empty; let also
$x^i,\xi_{1i},\xi_{2 \alpha}$ and ${y}^i,{\theta}_{1i},{\theta}_{2
  \alpha}$ be weakly canonical coordinates on $U_1$ and $U_2$,
respectively.  Then the functions ${\bar x}^i=x^i|_{\cL}$
(respectively, ${\bar y}^j=y^j|_{\cL}$) are local coordinates on
${\bar U}_1$ (respectively, ${\bar U}_2$).  We have seen that the
vector fields
\begin{equation}
  X_i=\abrkt{\xi_{1i}}{\cdot}^1=\ddl{}{x^i}\,, \qquad
  Y_j=\abrkt{\theta_{1j}}{\cdot}^1=\ddl{}{y^j}
\end{equation}
satisfy $\commut{X_i}{X_j}=0$ in $U_1$, $\commut{{Y}_i}{{Y}_j}=0$ in
$U_2$, and $\commut{X_i}{{Y}_j}=0$ in $U_1 \cap U_2$.  These vector
fields are tangent to $\cL$ and, thus, determine commuting vector
fields ${\bar X}_i=\ddl{}{{\bar x}^i}$ and ${{\bar Y}}_i=\ddl{}{{\bar
    y}^i}$ on ${\bar U}_1$ and ${\bar U}_2$, respectively.  In ${\bar
  U}_1\cap {\bar U}_2$, we have
\begin{equation}
  0=\commut{ {\bar X}_i  }{ {{\bar Y}}_j  }=
  \ddl{}{{\bar x}^i} {{\bar Y}}_j^k \ddl{}{{\bar x}^k}\,,
\end{equation}
where ${{\bar Y}}_j^k={{\bar Y}}_j.{\bar x}^k=\ddl{{\bar x}^k}{{\bar
    y}^j}$ are the coefficients of the vector field ${\bar Y}_j$ in
the coordinates ${\bar x}^i$. This means that ${\bar Y}^k_j$, which is
the Jacobi matrix associated with the change of coordinates ${\bar x}
\to {{\bar y}}$, is a constant matrix.  Thus, there exists an atlas on
$\cL$ such that the Jacobi matrices are constant. This is equivalent
to the statement of the theorem.

\medskip

This raises the question as to the geometric structures on~$\cL$ that
give rise to a pair of compatible antibrackets on a vector bundle
over~$\cL$. Recall that in the $Sp(2)$-symmetric quantization, the
manifold $\cL$ is the space that includes the original fields of the
theory to be quantized, the ghosts, antighosts, and the auxiliary
fields in the $Sp(2)$-symmetric quantization \cite{[BLT]}, which in
the triplectic case include also the `symplectic'
partners~\cite{[BM0],[BMS],[BM]}. One then adds antifields, thereby
constructing the triplectic manifold~$\cM$.  In the case where $\cL$
is a linear (super)space, the known
construction~\cite{[BLT],[BM0],[BMS]} works by assigning each field
$\phi^A$ (a coordinate on $\cL$) a pair of antifields $\phi^*_{aA}$.
Then the nonvanishing antibrackets read as
$\abrkt{\phi^A}{\phi^*_{Bb}}^a=\delta^a_b \delta^A_B$; these
antibrackets are evidently covariant under the linear transformation
of $\cL$ combined with the induced transformations of $\phi^*_{aA}$.
When $\cL$ is not a linear (super)space, we see that it cannot be
arbitrary: it has to admit a {\it flat\/} connection. Once the
connection is given, we can consider the `duplicated' cotangent bundle
$\cM=\Pi T^*\cL\oplus\Pi T^*\cL$ over $\cL$ with the reversed parity
of the fibers.
In some coordinate neighbourhood on $\cL$, the coordinates on the
fibers of $\Pi T^*\cL\oplus\Pi T^*\cL$ read as $\xi_{ai}$, $a=1,2$,
$\e{\xi_{ai}}=\e{x^i}+1$.  Under coordinate changes on $\cL$, the
coordinates $\xi_{ai}$ transform as $\ddl{}{x^i}$.  We now construct
the antibrackets as
\begin{equation}
  \abrkt{x^i}{\xi_{bj}}^a=\delta^a_b \delta^i_j\,, \quad
  \abrkt{\xi_{ai}}{\xi_{bj}}^c=
  \delta^c_b \Gamma^m_{ij} \xi_{am}-
  \delta^c_a \Gamma^m_{ij} \xi_{bm}\,,
  \label{Form-with-Gamma}
\end{equation}
where $\Gamma^m_{ij}$ are the Christoffel coefficients of $\nabla$ in
the coordinate system $x$ on $\cL$.  Introducing ${\bar \nabla}=
\ddl{}{x^i}+\Gamma^{m}_{ij}\xi_{bm}\ddl{}{\xi_{bj}}$, we can
rewrite~\req{Form-with-Gamma} as
\begin{equation}
  \abrkt{F}{G}^a=-G \ddr{}{\xi_{ai}} {\bar\nabla}_i F+\sign{F}{G}
  F \ddr{}{\xi_{ai}} {\bar\nabla}_i G\,.
\end{equation}
The symmetrized Jacobi identities for this pair of antibrackets are
satisfied because the curvature and torsion of $\nabla$ vanish.

Thus, we have seen that in triplectic quantization, the manifold of
fields $\phi^A$ is required to admit a flat symmetric connection.
This is in contrast with the standard BV-scheme, where no additional
requirements are imposed on the manifold of {\it fields\/}.  This may
be viewed as a restriction on the applicability of the covariant
$Sp(2)$ quantization.

\section{Conclusions}
We have uncovered the geometric structures underlying the triplectic
quantization of gauge theories. The most essential of these is the
K\"ahler manifold with additional polarizations (a certain analogue of
a hyper-K\"ahler manifold, however with a ``wrong'' signature of two
complex structures).

As we have seen, however, the requirements on the marked functions
from~\cite{[GS]} that lead eventually to a Darboux-like theorem
restrict the spaces involved in the quantization by a number of
flatness conditions
This may be viewed as a limitation of the entire $Sp(2)$-symmetric
quantization approach; alternatively, it can be attributed to the
properties of the axioms imposed in~\cite{[GS]}.  Thus, one may
speculate that if the mutual commutativity condition imposed on the
antibrackets is relaxed, one may still be able to identify some
interesting geometries; the key question would then be about the
meaning of the quantization procedure (e.g., in the construction of
path integral).\footnote{We thank I.~Batalin for a discussion on this
  point.}

As a final remark, note that the geometric structures that we have
identified in the triplectic quantization (the symplectic and complex
structures and the transversal polarizations) are those entering the
Geometric Quantization (see, e.g.,~\cite{[GuiSt]}) of symplectic
manifolds. One may also note some formal similarities with the
structures discussed in~\cite{[Bering]} in the context of BV geometry.

\paragraph{Acknowledgements.} We wish to thank I.~A.~Batalin and
I.~V.~Tyutin for illuminating discussions. This work was supported in
part by the RFBR Grant~96-01-00482 and by the INTAS-RFBR
Grant~95-0829.

\def\theequation{A.\arabic{equation}}
\section*{Appendix}
In this paper, we make use of some conventions of differential
geometric objects on the supermanifold~$\cM$.  The main object is the
graded associative algebra $\func\cM$ of (globally defined) smooth
functions on $\cM$.  A vector field $X$ on $\cM$ is the differential of
$\func\cM$, which means that
\begin{equation}
  X(FG)=(XF)G-\sig{X}{F}F(XG)\,,
\end{equation}
where $\e{F}$ is the Grassmann parity of a function $F$.  The
Grassmann parity $\e{X}$ of a vector field $X$ is defined as
$\e{X}=\e{XF}+\e{F}\,,F\in\func\cM$.  Vector fields on $\cM$
constitute a left module over $\func\cM$; for any $F\in\func\cM$ and
any vector field $X$, we define $(FX)$ by its action on arbitrary $G
\in \func\cM$ as $(FX)G=F(XG)$.

An $N$-form $\Phi$ is defined as a multilinear mapping
$\Phi:\Vect_\cM\times\ldots\Vect_\cM\to\func\cM$ satisfying
\begin{equation}
  \begin{array}{c} \Phi(X_1,\ldots ,X_N)=
    \sign{X_j}{X_{j+1}}\Phi(X_1,\ldots,X_{j-1},X_{j+1},X_{j},X_{j+2},\ldots,X_N)\,,\\
    \Phi(FX_1,\ldots ,X_N)=F\Phi(X_1,\ldots ,X_N)\,,\quad X_1,\ldots,X_N
    \in \Vect_\cM\,,\quad F\in \func\cM\,.
  \end{array}
\end{equation}
The Grassmann parity $\e{\Phi}$ of the $N$-form $\Phi$ is defined as
\begin{equation}
  \e{\Phi}=\e{\Phi(X_1,\ldots,X_N)}+\e{X_1}+\ldots+\e{X_N}\,.
\end{equation}
The differential forms are a right module over $\func\cM$.  For any
$F\in \func\cM$ and N-form $\Phi$ we have
\begin{equation}
(\Phi F)(X_1,\ldots ,X_N)=\Phi(X_1,\ldots ,X_N)F\,.
\end{equation}
Contraction of the vector field $X$
and an $N$-form $\Phi$ reads as
\begin{equation}
  (\contraction_{X}\Phi)(X_1,\ldots,X_{N-1})=
  {p(\Phi)}\Phi(X_1,\ldots,X_{N-1},X)\,.
\end{equation}
where $p(\Phi)$ is the degree of the form~$\Phi$.  It is useful to
define the outer product $\wedge$ of forms such that $\contraction_X$
differentiate the outer product,
\begin{equation}
  \contraction_X (\Phi\wedge\Psi) =
  (\contraction_X \Phi)\wedge\Psi+
  (-1)^{(\e{X}+1)(p(\Phi) + \e{\Phi})}\Phi\wedge(\contraction_X\Psi)\,,
\end{equation}
Now the De Rham differential $d$ is by definition the nilpotent linear
operator satisfying
\begin{equation}
  \new\begin{array}{c}
    \contraction_X dF=\langle X,dF\rangle =XF\,,\quad X\in
    \Vect_\cM\,, \quad F\in \func\cM\,,\\
    d(\Phi \wedge \Psi)=(d\Phi)\wedge\Psi+
    (-1)^{p(\Phi)+\e{\Phi}}\Phi \wedge (d \Psi)\,,
  \end{array}
\end{equation}
In particular, for the 1-form $\Phi$ we have
\begin{equation}
  (d\Phi)(X_1,X_2)=
  \half(X_1\Phi(X_2)-\sig{X_1}{X_2}X_2\Phi(X_1)-\Phi(\commut{X_1}{X_2}))
  (-1)^{\e{X_2}+1} \,,
\end{equation}
where $X_1,X_2 \in \Vect_\cM \,$.

\end{document}